\documentclass[fleqn]{article}
\usepackage{longtable}
\usepackage{graphicx}

\begin {document}
\begin{flushleft}
{\LARGE
{\bf Electron Impact Excitation of O~III: An Assessment}
}\\

\vspace{1.5 cm}

{\bf {Kanti  M  ~Aggarwal}}\\ 

\vspace*{1.0cm}

Astrophysics Research Centre, School of Mathematics and Physics, Queen's University Belfast, \\Belfast BT7 1NN, Northern Ireland, UK\\ 
\vspace*{0.5 cm} 

e-mail: K.Aggarwal@qub.ac.uk \\

\vspace*{0.20cm}


\vspace*{1.0 cm}

\vspace*{1.0 cm}
\vspace*{1.0 cm}

\hrule

\vspace{0.5 cm}

\end{flushleft}


\vspace*{1.0 cm}

\begin{abstract}

 Tayal and  Zatsarinny [{\em Astrophys. J.}    {\bf 2017}, {\emph 850},   147] have reported results for  energy levels, radiative rates (A-values), lifetimes,  and effective collision strengths ($\Upsilon$) for transitions among 202 levels of C-like O~III. For the calculations they have  adopted the multi-configuration Hartree-Fock (MCHF) code for the energy levels and A-values, and B-spline $R$-matrix (BSR) code for $\Upsilon$. Their reported results cover a (much) larger range of levels/transitions than generally available in the literature, and appear to be accurate for energy levels and A-values. However, the magnitude and behaviour  of $\Upsilon$ do not appear to be correct for several transitions. We demonstrate this  through our independent calculations by adopting the flexible atomic code (FAC) and recommend a fresh calculation for this important ion. 
\end{abstract}

\clearpage

\section{Introduction}\label{sec1}

C-like O~III is an ion of astrophysical importance, because many of its emission lines are frequently observed in a variety of plasmas, such as planetary nebulae, H~II regions and solar atmospheres, and have been very useful for both temperature and density diagnostics  -- see for example  the paper by Tayal and  Zatsarinny  \cite{sst} and references therein. For this reason, there have been several calculations for the generation of its atomic data, mainly for energy levels, radiative rates (A-values), collision strengths  ($\Omega$),  and effective collision strengths ($\Upsilon$). Measurements for some energy levels have been performed by a few workers and these have been compiled and  assessed  by the NIST (National Institute of Standards and Technology) team, and their recommended values are freely available at their website:   {\tt http://www.nist.gov/pml/data/asd.cfm}. However, for the remaining parameters one has to depend on theoretical results. 

During the past four decades there have been several calculations for the determination of A-values, $\Omega$ and $\Upsilon$, for which a variety of methods have been adopted. Particularly for the calculations of $\Omega$ and $\Upsilon$, required for diagnostics and modelling of plasmas, the two most common methods are {\em distorted wave} (DW) and $R$-matrix. The main difference between the two is in the exclusion/inclusion of closed-channel (Feshbach) resonances, which have significant effect on the values of $\Upsilon$, particularly for the forbidden transitions, and at almost all temperatures of interest. For this reason, the $R$-matrix calculations are preferred for (mainly) $\Upsilon$. 

The first $R$-matrix calculations for O~III were initiated by Baluja et al.  \cite{bbk}. After that many other have followed, but the most important and widely used one are those of  Aggarwal and Keenan \cite{oiii}, who reported results for transitions among 46 levels belonging to the 2s$^2$2p$^2$, 2s2p$^3$, 2p$^4$, and 2s$^2$2p3$\ell$ configurations. This range of levels/transitions was considerably extended by  Storey et al. \cite{ssb}, who considered 146 levels. Unfortunately, their results have (very) limited application because their focus was only on 10 transitions belonging to the 5 levels of the 2s$^2$2p$^2$ ground configuration alone.  Therefore,    Tayal and  Zatsarinny \cite{sst} have performed yet another calculation for this important ion, and have considered 202 levels of the 2s$^2$2p$^2$, 2s2p$^3$, 2p$^4$, 2s$^2$2p3$\ell$,  2s$^2$2p4$\ell$, 2s2p$^2$3$\ell$, and 2s$^2$2p5s configurations, 14 in total. 

 For the generation of wavefunctions, i.e. for the calculations of energy levels and A-values, Tayal and  Zatsarinny  \cite{sst}  have adopted  the multi-configuration Hartree-Fock (MCHF) code (in combination with the B-spline expansion), and for the collisional data, i.e. $\Omega$ and $\Upsilon$, the B-spline $R$-matrix (BSR) code.  Since they have used flexible term dependent orbital sets to represent the target states, their calculated energies and A-values are (apparently) accurate, as shown in their tables 1 and 2. Regarding the other parameters, namely $\Omega$ and $\Upsilon$, they have presented results for only the latter, but for {\em all} transitions (20~301) among the considered 202 levels. Since corresponding data for 1035 transitions among the lowest 46 levels are already available \cite{oiii}, they have assessed the accuracy of their results based on comparisons for these, but have made no attempt to assess similar accuracy for the remaining 19~266 (95\%) transitions. Therefore, our main focus is on these transitions among the higher lying levels to see if these results are equally accurate.  
 
In the absence of corresponding $\Omega$ data it is (comparatively) more difficult to make accuracy assessments, particularly when the oscillator strengths (f-values) or A-values are also not given. Nevertheless, we make an attempt by other means, mainly by performing our own calculations with the easily available and generally reliable {\em Flexible Atomic Code} (FAC) of Gu  \cite{fac}, currently hosted at the website:  {\tt https://www-amdis.iaea.org/FAC/}. This code is based on the DW method, is fully relativistic, is highly efficient to run, and more importantly, produces results for background values of collision strengths ($\Omega_B$) which are comparable with other methods, such as the $R$-matrix.  This has been demonstrated in many of our earlier papers for a wide range of ions -- see for example, figure~2 of Aggarwal and Keenan  \cite{mgv} for Mg~V. However, resonances are (generally) not resolved in this code, but results for $\Omega_B$ will be helpful to draw conclusions.  

Before we discuss the results of Tayal and  Zatsarinny  \cite{sst} in detail, we will like to note that in a more recent paper Mao et. al. \cite{icft} have reported $\Upsilon$ data for several C-like ions (N~II to Kr~XXXI), including O~III, and for even a larger number of levels, i.e. 590 which arise from 24 configurations of 2{$\ell{^3}{n}\ell'$ with $n$ = 2 to 4, $\ell$ = 0 and 1, and $\ell'$ = 0 to 3, {\em plus} three additional configurations of 2s$^2$2p5$\ell$ with $\ell$ = 0, 1 and 2. For the calculations they have used the {\em AutoStructure} (AS) code to generate the wavefunctions, i.e. to calculate energy levels and the A-values, and for the values of $\Omega$ and subsequently $\Upsilon$ their {\em intermediate coupling frame transformation} (ICFT) method alongwith the standard $R$-matrix code. Although these calculations are the {\em largest} performed so far, and are the most recent one, yet their accuracy and the subsequent use for applications, particularly for the analysis of the nebular plasmas,  has recently been questioned by Morisset et al. \cite{atoms}, who have assessed their data to be less accurate, especially at low temperatures, and have actually warned against their use.  Since these calculations are the largest, as already stated, no direct comparisons are possible with other existing results, and therefore a detailed  assessment of the accuracy of their data, for a larger number of transitions and over a wider range of  temperature, may be the subject of a separate paper.  Furthermore, the deficiency of the ICFT calculations has already been discussed in the literature for several other ions \cite{mg5}, although in the  concerned calculations the authors  have made some improvements, particularly by extending the energy range of the $\Omega$ data up to three times the ionization potential, before extrapolating the range, for the determination of $\Upsilon$ up to very high temperatures, required for fusion plasmas.

\begin{figure*}
\setcounter{figure}{0}
\includegraphics[angle=-90,width=0.9\textwidth]{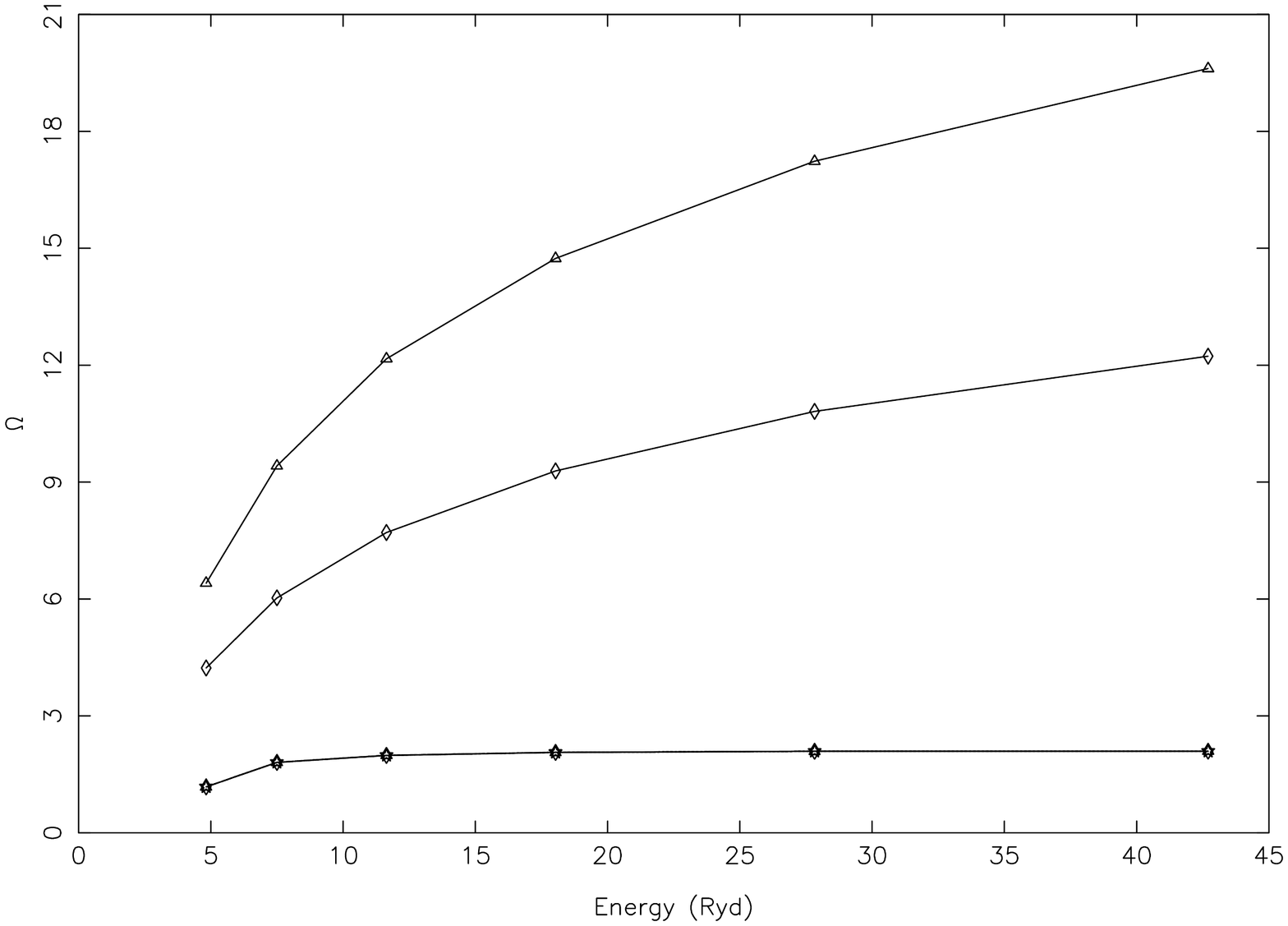}
 \vspace{-1.5cm}
\caption{Our calculated values of  $\Omega$  with FAC for  the 44--202 (triangles: 2s$^2$2p($^2$P$^o$)3d~$^1$F$^o_3$--2s2p$^2$($^2$P)3d~$^1$D$_2$), 45--202 (diamonds: 2s$^2$2p($^2$P$^o$)3d~$^1$P$^o_1$--2s2p$^2$($^2$P)3d~$^1$D$_2$), and 164--202 (stars: 2s2p$^2$($^2$P)3s~$^1$P$_1$--2s2p$^2$($^2$P)3d~$^1$D$_2$) transitions of O~III. }\label{f1}
 \end{figure*}

\begin{figure*}
\setcounter{figure}{1}
\includegraphics[angle=-90,width=0.9\textwidth]{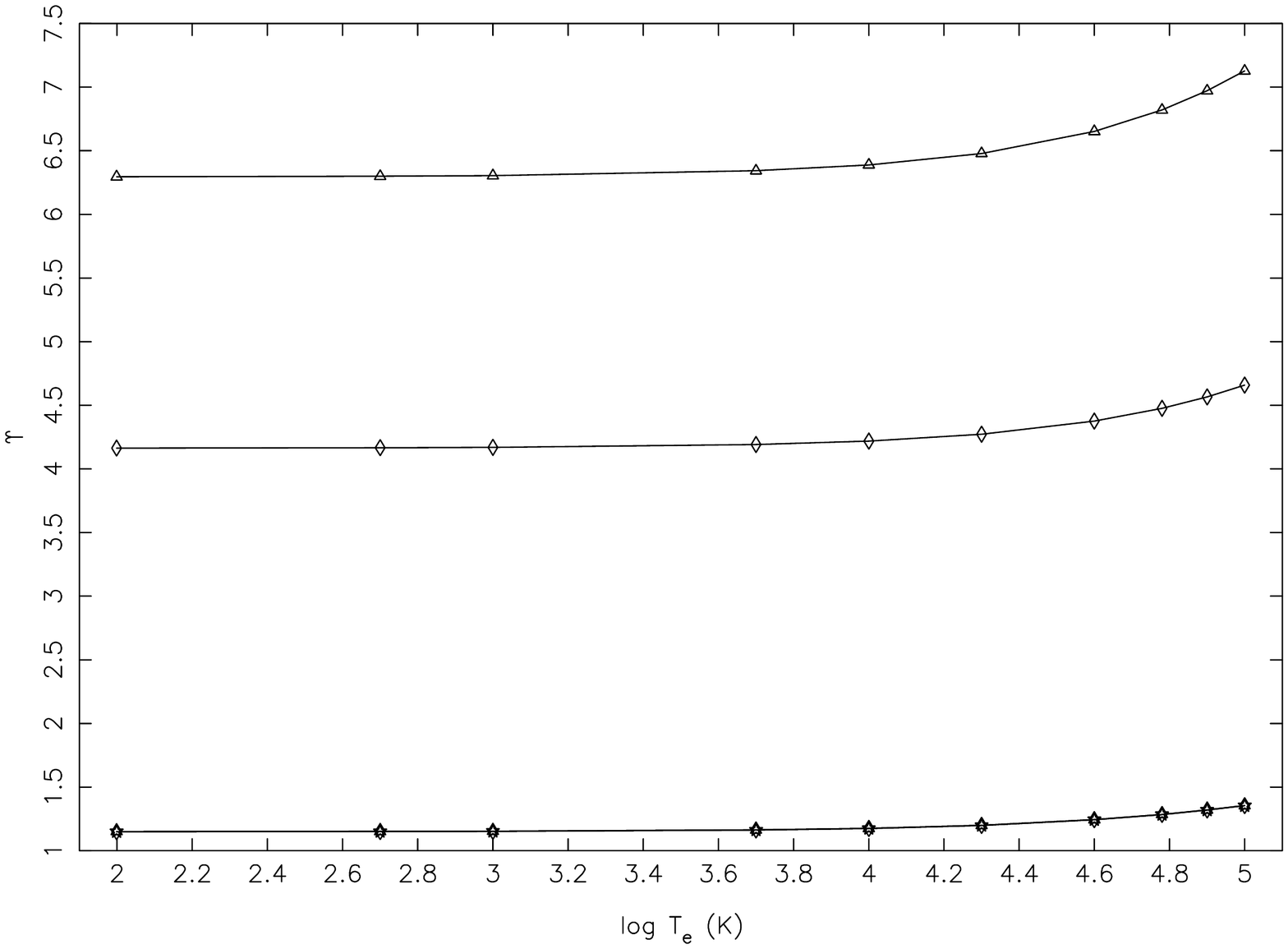}
 \vspace{-1.5cm}
\caption{Our calculated values of  $\Upsilon$  with FAC for  the 44--202 (triangles: 2s$^2$2p($^2$P$^o$)3d~$^1$F$^o_3$--2s2p$^2$($^2$P)3d~$^1$D$_2$), 45--202 (diamonds: 2s$^2$2p($^2$P$^o$)3d~$^1$P$^o_1$--2s2p$^2$($^2$P)3d~$^1$D$_2$), and 164--202 (stars: 2s2p$^2$($^2$P)3s~$^1$P$_1$--2s2p$^2$($^2$P)3d~$^1$D$_2$) transitions of O~III. } \label{f2}
\end{figure*}

\begin{figure*}
\setcounter{figure}{2}
\includegraphics[angle=-90,width=0.9\textwidth]{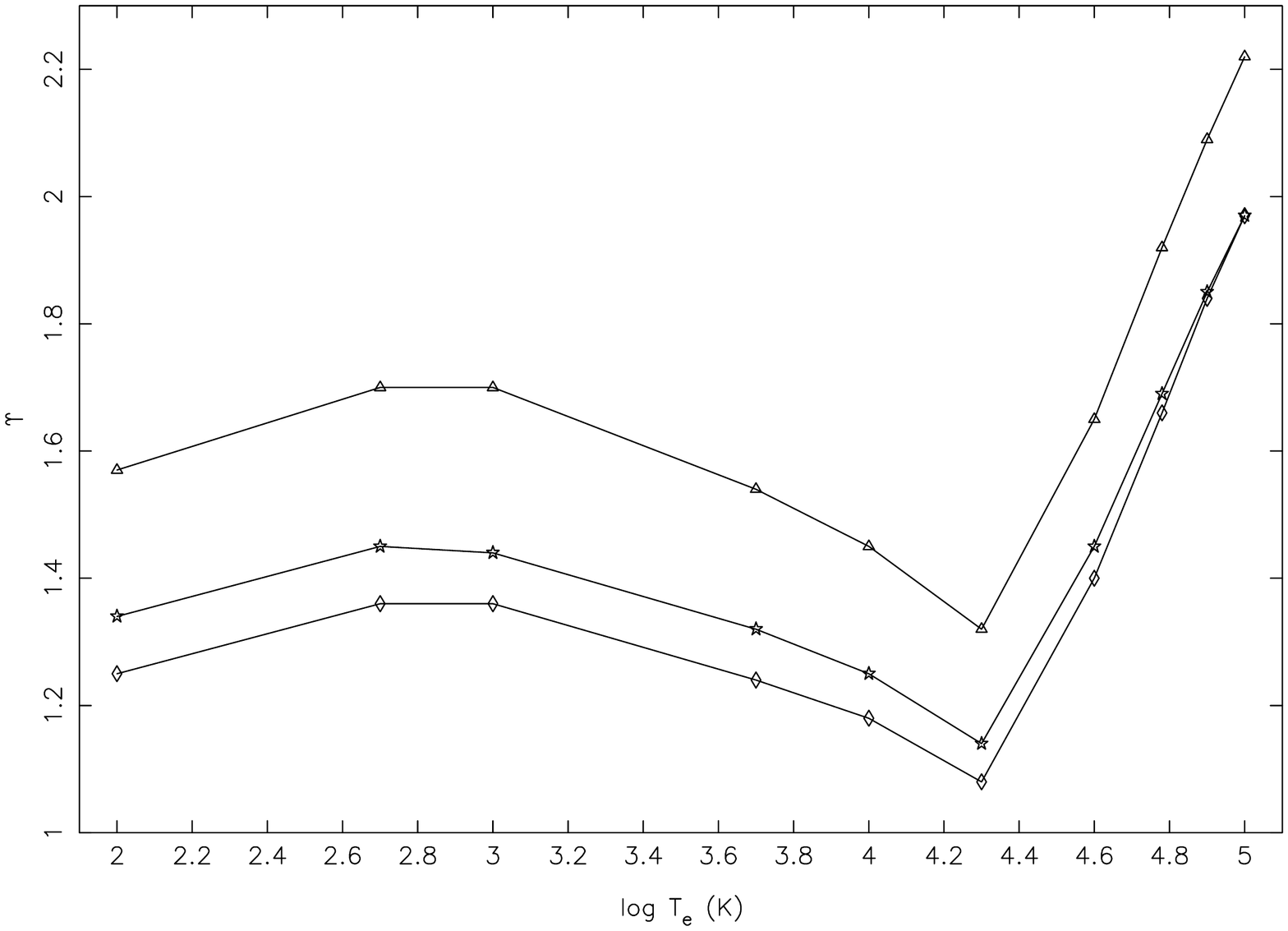}
 \vspace{-1.5cm}
 \caption{The calculated values of   $\Upsilon$ by Tayal and Zatsarinny \cite{sst} with BSR for  the 44--202 (triangles: 2s$^2$2p($^2$P$^o$)3d~$^1$F$^o_3$--2s2p$^2$($^2$P)3d~$^1$D$_2$), 45--202 (diamonds: 2s$^2$2p($^2$P$^o$)3d~$^1$P$^o_1$--2s2p$^2$($^2$P)3d~$^1$D$_2$), and 164--202 (stars: 2s2p$^2$($^2$P)3s~$^1$P$_1$--2s2p$^2$($^2$P)3d~$^1$D$_2$) transitions of O~III. }\label{f3}
 \end{figure*}
 
 \section {Collision strengths and effective collision strengths}\label{sec2}

For our calculations we have considered the same 202 levels among the 14 configurations (given in section~\ref{sec1}) as by  Tayal and  Zatsarinny  \cite{sst}, and exclusively listed in their table~1. For easy understanding we will use the same level {\em indices}, while referring to a transition, as by them. In Figure~\ref{f1} we show our results for $\Omega$ for three  transitions, namely 44--202 (2s$^2$2p($^2$P$^o$)3d~$^1$F$^o_3$--2s2p$^2$($^2$P)3d~$^1$D$_2$, f = 0.2798), 45--202 (2s$^2$2p($^2$P$^o$)3d~$^1$P$^o_1$--2s2p$^2$($^2$P)3d~$^1$D$_2$, f = 0.3960), and 164--202 (2s2p$^2$($^2$P)3s~$^1$P$_1$--2s2p$^2$($^2$P)3d~$^1$D$_2$), out of which the first two are allowed and have significant f-values,  calculated with FAC, whereas the third one is parity forbidden,  and all energies shown in this (and other) figures are {\em incident}. Since all these three (and many more) transitions involve the highest level of the calculations, their values of $\Omega$ (in any calculation) vary smoothly, as expected, although no comparisons can be made with the corresponding (non available) data of  Tayal and  Zatsarinny   \cite{sst}. Our calculated values of $\Upsilon$ are shown in Figure~\ref{f2} in the same temperature range as by them, i.e. between 100 and 100~000~K, and the values (expectedly) vary smoothly with increasing T$_e$. The corresponding results of Tayal and Zatsarinny are shown in Figure~\ref{f3},   the magnitudes of which  are not only {\em lower}, by up to a factor of four, especially for the allowed transitions, but their behaviour are also anomalous, particularly towards the higher end of the temperature range. Some differences in $\Upsilon$ values may be due to the corresponding differences in the f-values, but since both 44--202 and 45--202 are strong transitions, their f-values are not expected to differ by more than 20\% -- see for example, similar calculations for S~III in table~1 of  Aggarwal \cite{s3}.  Before we discuss the (possible) reasons for the discrepancies we consider a few more transitions to show that the above three are not the isolated examples. 

In Figure~\ref{f4} we show our $\Omega$ for three other transitions, namely 185--202 (2s2p$^2$($^2$P)3p~$^1$D$^o_2$--2s2p$^2$($^2$P)3d~$^1$D$_2$, f = 0.0896), 186--202 (2s2p$^2$($^2$P)3p~$^1$P$^o_1$--2s2p$^2$($^2$P)3d~$^1$D$_2$, f = 0.2899), and 201--202 (2s2p$^2$($^2$P)3d~$^1$F$_3$--2s2p$^2$($^2$P)3d~$^1$D$_2$). Again, the first two are allowed and the third one is forbidden, but the magnitudes of all these three transitions are significantly larger than of those shown in Figure~\ref{f1}.  This is because these transitions have very small energies ($<$ 4~eV) and hence converge slowly. Our results of $\Upsilon$ and those of   Tayal and  Zatsarinny \cite{sst} are shown in Figures~\ref{f5}-\ref{f6}. Unfortunately the trends (and hence the conclusions) are the same as for those shown in Figure~\ref{f1}, i.e. their magnitudes are lower than those of ours by up to a factor of four and the behaviours are incorrect.

The lower magnitudes of $\Upsilon$ of   Tayal and  Zatsarinny \cite{sst} are (probably) due to the insufficient number of partial waves included by them, i.e. with angular momentum $J \le$ 25.5. Although they have included the contribution of higher neglected partial waves through a top-up procedure, it still underestimates the magnitudes if implemented at an early stage, and an example of this can be seen in  figure~2 of  Aggarwal and Keenan  \cite{nixi} for a transition of Ni~XI. For the incorrect behaviours seen in Figures~\ref{f3} and \ref{f6}, there are three possibilities. First, there may be pseudo-resonances, at energies above thresholds, which abnormally affect the determinations of $\Upsilon$ results. Second, their adopted energy mesh of 0.2~Ryd at energies above thresholds, is too coarse, in comparison to the very fine 0.000~04~Ryd in the thresholds region, i.e. up to about 4.5~Ryd. However, the possibility of pseudo-resonances is less likely  because of the nature of the wavefunctions they adopted. Otherwise also, their presence (if any) will increase the magnitude of $\Upsilon$, rather than decreasing its value.  Moreover, they have also mentioned a smooth variation of $\Omega$ at higher energies, and have demonstrated that in their figures~1--3.  Therefore, (one of) the definite reason for the anomalous behaviour of their $\Upsilon$ results is the ``coarse" energy mesh adopted at higher energies. This particularly affects transitions involving the higher levels, as already noted in Figures~\ref{f3} and \ref{f6}, and also discussed earlier for transitions of Mg~V   \cite{mgv} and S~III \cite{s3}.  Finally, the third possibility is the {\em extrapolation} they performed for determining $\Omega$ values beyond 30~Ryd, the energy up to which they did actual calculations. Often the formulae adopted for the extrapolations (for various types of transitions) are not very robust and many examples of this have been discussed in our earlier work \cite{mg5}. As a result of this sometimes the extrapolations do more harm than good and it is clear from the behaviour of $\Upsilon$ shown in Figures~\ref{f3} and \ref{f6}. More importantly, it is beyond our understanding why they had to do the extrapolation in the first place, because the highest threshold is below 5~Ryd (see their table~1) and therefore they have a wide range of over 25~Ryd to calculate $\Upsilon$ up to 10$^5$~K, which is only 0.63~Ryd. Hence, the anomalous behaviours of their $\Upsilon$ results are (probably) due to a combination of the coarse energy mesh as well as the (unnecessary) extrapolations. 

\begin{figure*}
\setcounter{figure}{3}
\includegraphics[angle=-90,width=0.9\textwidth]{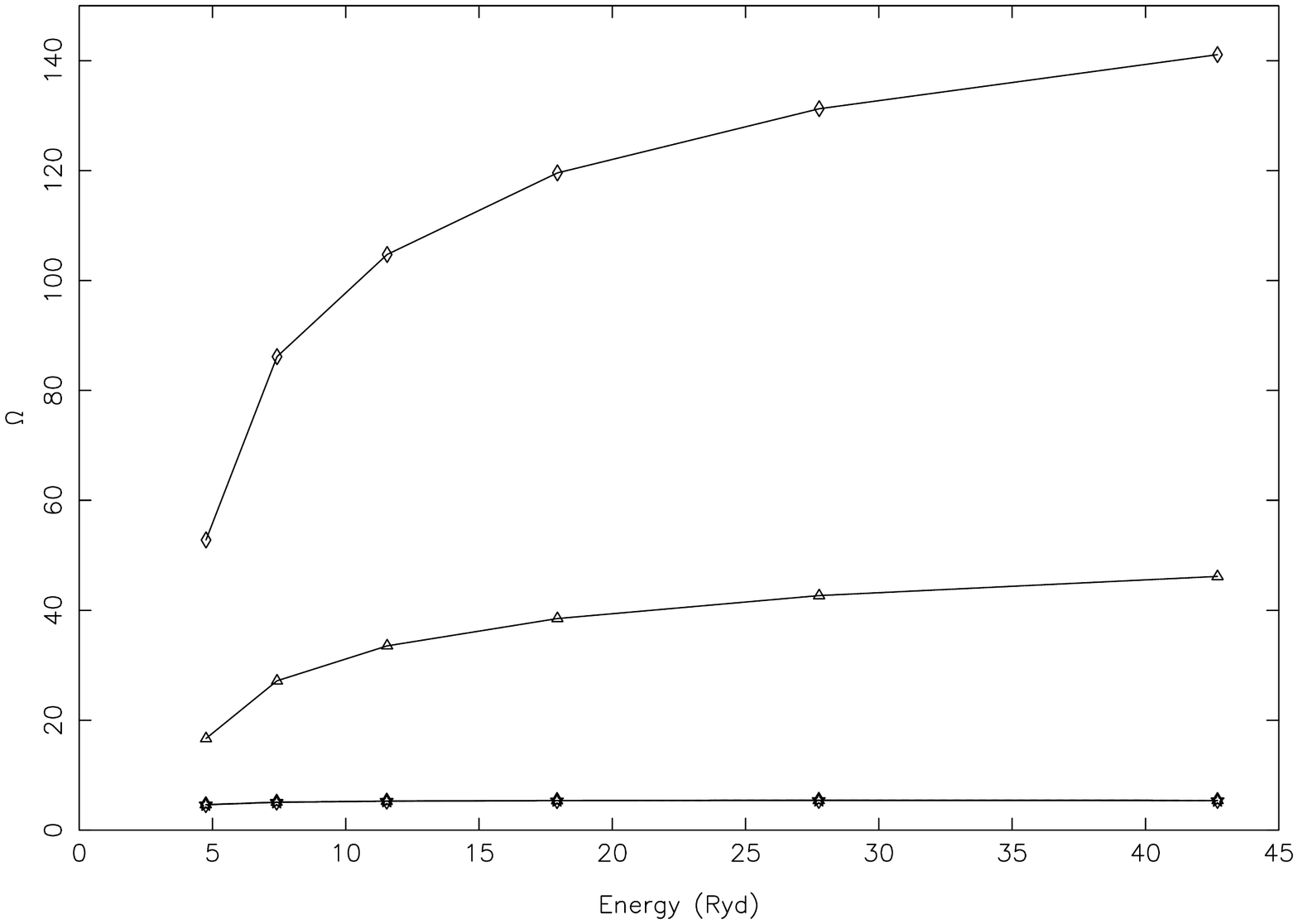}
 \vspace{-1.5cm}
 \caption{Our calculated values of  $\Omega$   with FAC for  the 185--202 (triangles: 2s2p$^2$($^2$P)3p~$^1$D$^o_2$--2s2p$^2$($^2$P)3d~$^1$D$_2$), 186--202 (diamonds: 2s2p$^2$($^2$P)3p~$^1$P$^o_1$--2s2p$^2$($^2$P)3d~$^1$D$_2$), and 201--202 (stars: 2s2p$^2$($^2$P)3d~$^1$F$_3$--2s2p$^2$($^2$P)3d~$^1$D$_2$)  transitions of O~III. }\label{f4}
 \end{figure*}

\begin{figure*}
\setcounter{figure}{4}
\includegraphics[angle=-90,width=0.9\textwidth]{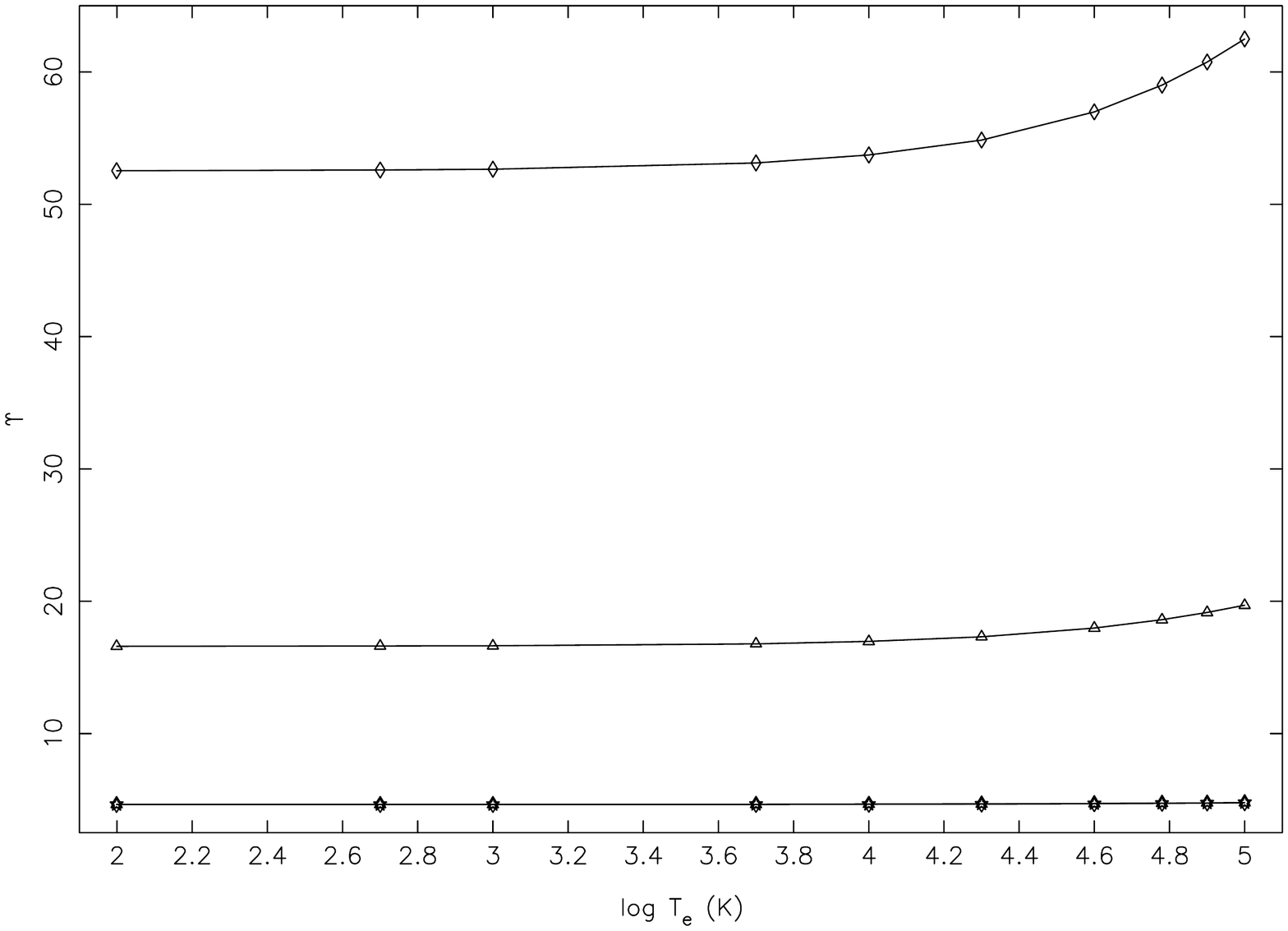}
 \vspace{-1.5cm}
 \caption{Our calculated values of  $\Upsilon$  with FAC for  the 185--202 (triangles: 2s2p$^2$($^2$P)3p~$^1$D$^o_2$--2s2p$^2$($^2$P)3d~$^1$D$_2$), 186--202 (diamonds: 2s2p$^2$($^2$P)3p~$^1$P$^o_1$--2s2p$^2$($^2$P)3d~$^1$D$_2$), and 201--202 (stars: 2s2p$^2$($^2$P)3d~$^1$F$_3$--2s2p$^2$($^2$P)3d~$^1$D$_2$)  transitions of O~III. }\label{f5}
 \end{figure*}

\begin{figure*}
\setcounter{figure}{5}
\includegraphics[angle=-90,width=0.9\textwidth]{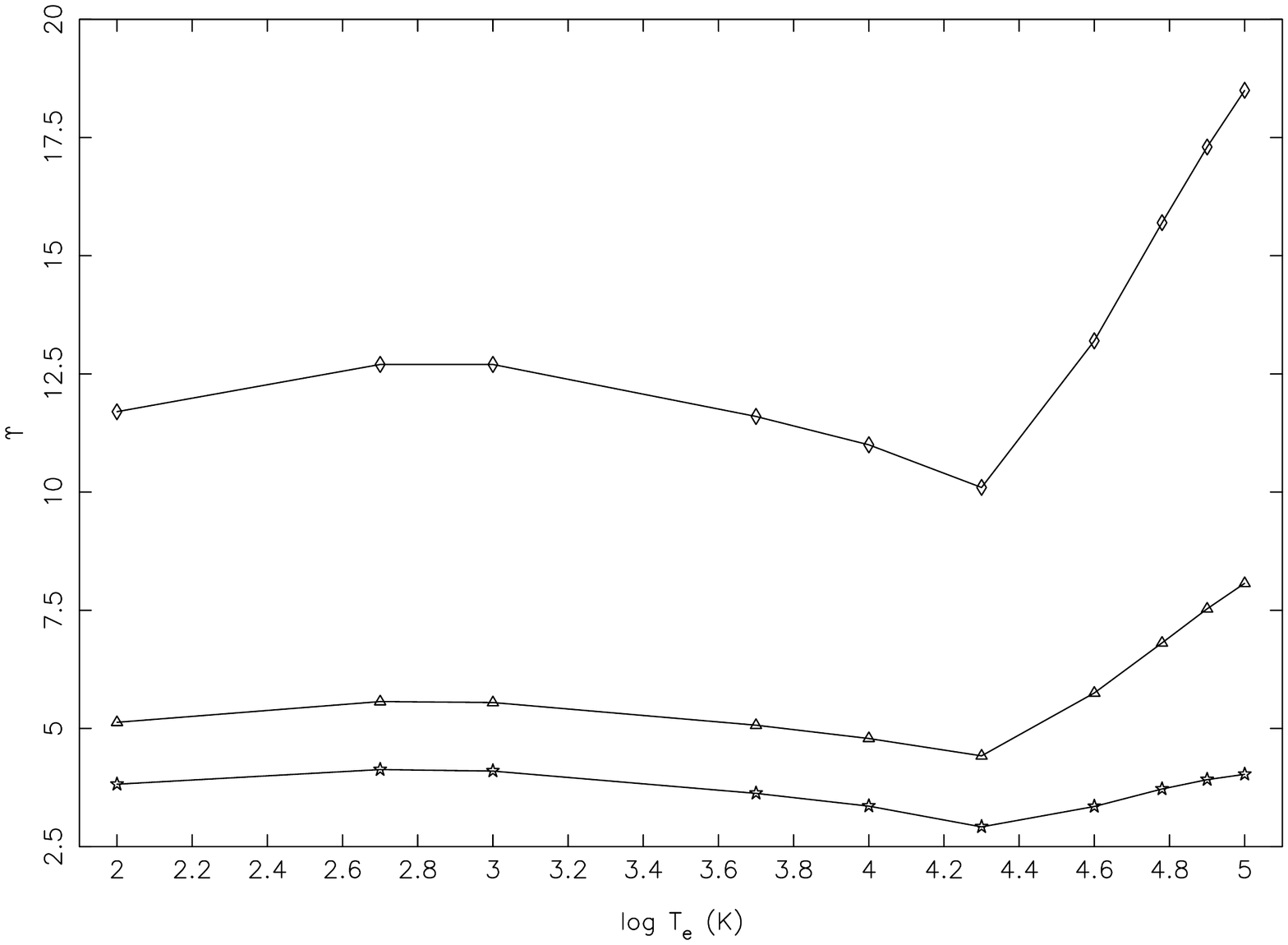}
 \vspace{-1.5cm}
 \caption{The calculated values of   $\Upsilon$  by Tayal and Zatsarinny \cite{sst} with BSR for  the 185--202 (triangles: 2s2p$^2$($^2$P)3p~$^1$D$^o_2$--2s2p$^2$($^2$P)3d~$^1$D$_2$), 186--202 (diamonds: 2s2p$^2$($^2$P)3p~$^1$P$^o_1$--2s2p$^2$($^2$P)3d~$^1$D$_2$), and 201--202 (stars: 2s2p$^2$($^2$P)3d~$^1$F$_3$--2s2p$^2$($^2$P)3d~$^1$D$_2$)  transitions of O~III. }\label{f6}
 \end{figure*}

\begin{figure*}
\setcounter{figure}{6}
\includegraphics[angle=-90,width=0.9\textwidth]{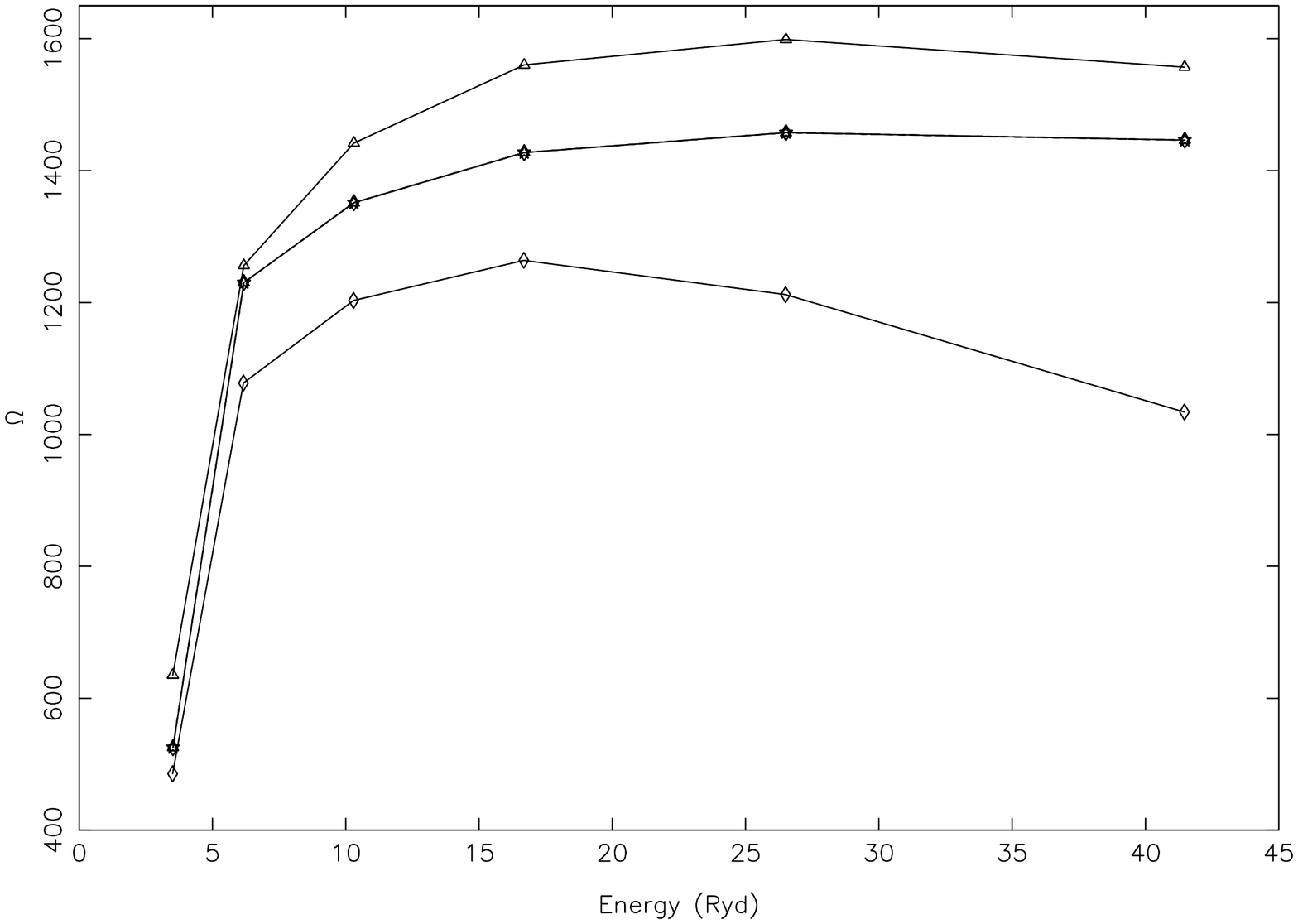}
 \vspace{-1.5cm}
 \caption{Our calculated values of $\Omega$   with FAC  for  the 83--98 (triangles: 2s$^2$2p($^2$P$^o$)4d~$^3$F$^o_4$--2s$^2$2p($^2$P$^o$)4f~$^3$G$_5$), 89--93 (diamonds: 2s$^2$2p($^2$P$^o$)4d~$^3$D$^o_3$--2s$^2$2p($^2$P$^o$)4f~$^3$F$_4$), and 94--101 (stars: 2s$^2$2p($^2$P$^o$)4d~$^1$F$^o_3$--2s$^2$2p($^2$P$^o$)4f~$^1$G$_4$) transitions of O~III. }\label{f7}

 \end{figure*}

\begin{figure*}
\setcounter{figure}{7}
\includegraphics[angle=-90,width=0.9\textwidth]{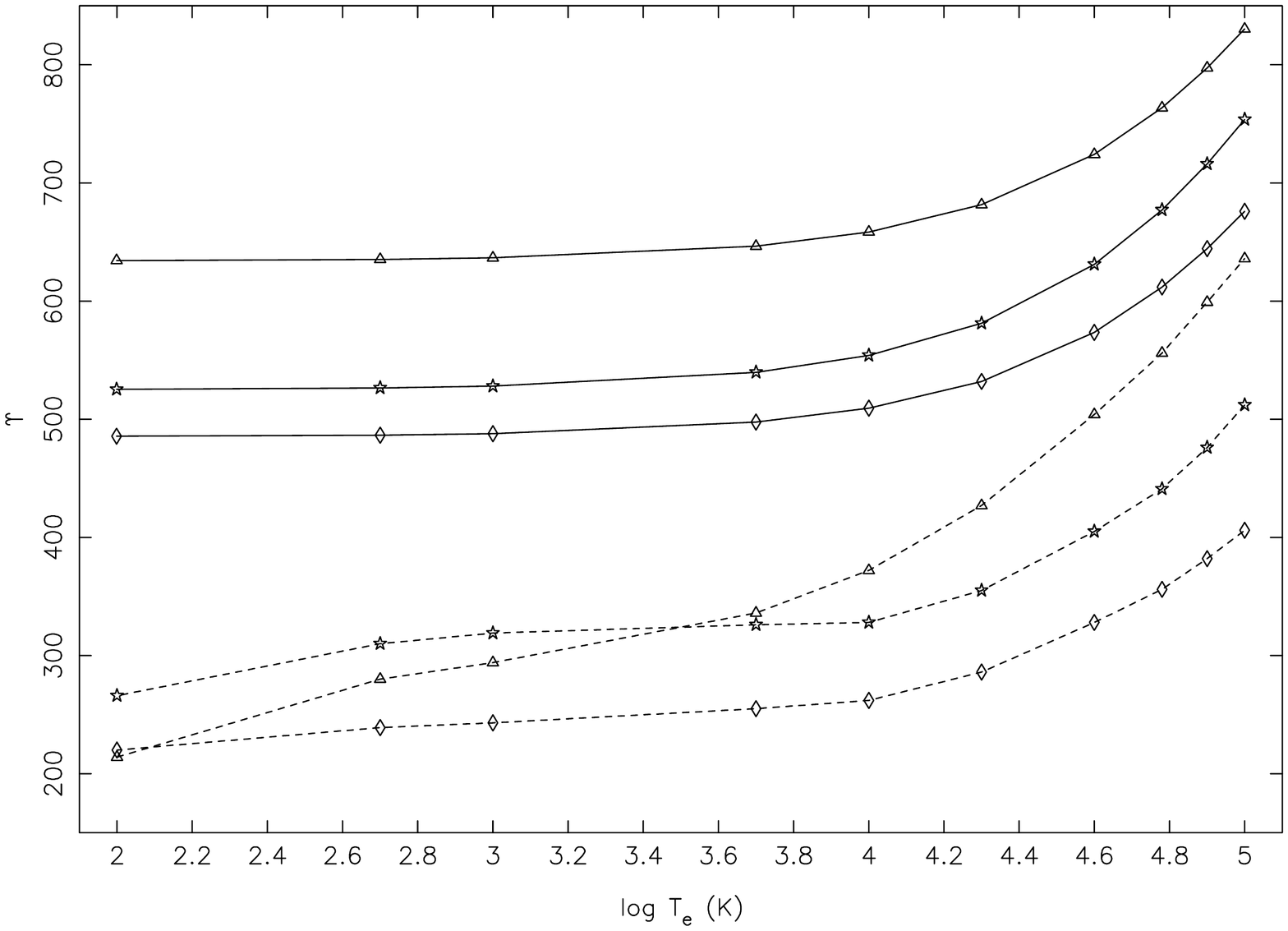}
 \vspace{-1.5cm}
 \caption{Our calculated values of  $\Upsilon$  with FAC  for  the 83--98 (triangles: 2s$^2$2p($^2$P$^o$)4d~$^3$F$^o_4$--2s$^2$2p($^2$P$^o$)4f~$^3$G$_5$), 89--93 (diamonds: 2s$^2$2p($^2$P$^o$)4d~$^3$D$^o_3$--2s$^2$2p($^2$P$^o$)4f~$^3$F$_4$), and 94--101 (stars: 2s$^2$2p($^2$P$^o$)4d~$^1$F$^o_3$--2s$^2$2p($^2$P$^o$)4f~$^1$G$_4$) transitions of O~III.  Continuous curves: present results with FAC and broken curves: earlier results of Tayal and Zatsarinny \cite{sst} with BSR.}\label{f8}
 \end{figure*}
 
 We now focus our attention on some other transitions which do not involve the higher lying levels. In Figure~\ref{f7} we show our values of $\Omega$ for three transitions, namely 83--98 (2s$^2$2p($^2$P$^o$)4d~$^3$F$^o_4$--2s$^2$2p($^2$P$^o$)4f~$^3$G$_5$), 89--93 (2s$^2$2p($^2$P$^o$)4d~$^3$D$^o_3$--2s$^2$2p($^2$P$^o$)4f~$^3$F$_4$), and 94--101 (2s$^2$2p($^2$P$^o$)4d~$^1$F$^o_3$--2s$^2$2p($^2$P$^o$)4f~$^1$G$_4$), all of which are allowed {\em and} have large magnitudes. The corresponding results of $\Upsilon$ from our calculations and those of Tayal and  Zatsarinny  \cite{sst} are shown in Figure~\ref{f8}. As earlier, their $\Upsilon$ values are underestimated by over a factor of three at most temperatures, although the differences decrease with increasing T$_e$. The reason for these large differences, particularly at the lower end of T$_e$, is rather simple. The thresholds region is often dominated by resonances (although mostly for forbidden transitions) and therefore a top-up for higher neglected partial waves is performed only at energies {\em above} thresholds. Subsequently, values of $\Omega$ in the thresholds region remain uncorrected if sufficient number of partial waves are not included, as is the case with their calculations. This underestimation can be confirmed by an example shown in figure~4 of  Aggarwal et al. \cite{fexv} for the 3s3p~$^1$P$^o_1$ -- 3p$^2$~$^1$D$_2$ transition of Fe~XV in which  the whole $\Omega_B$ is lifted upwards with the inclusion of larger ranges of $J$. The threshold region for these transitions is of over 1.2~Ryd, about twice the range of T$_e$ (0.63~Ryd) considered by them, and therefore values of $\Upsilon$ are underestimated over the entire T$_e$ range. Since these (and several other) transitions have comparatively larger magnitudes their effect on modelling applications are larger than of the weaker ones.

\begin{figure*}
\setcounter{figure}{8}
\includegraphics[angle=-90,width=0.9\textwidth]{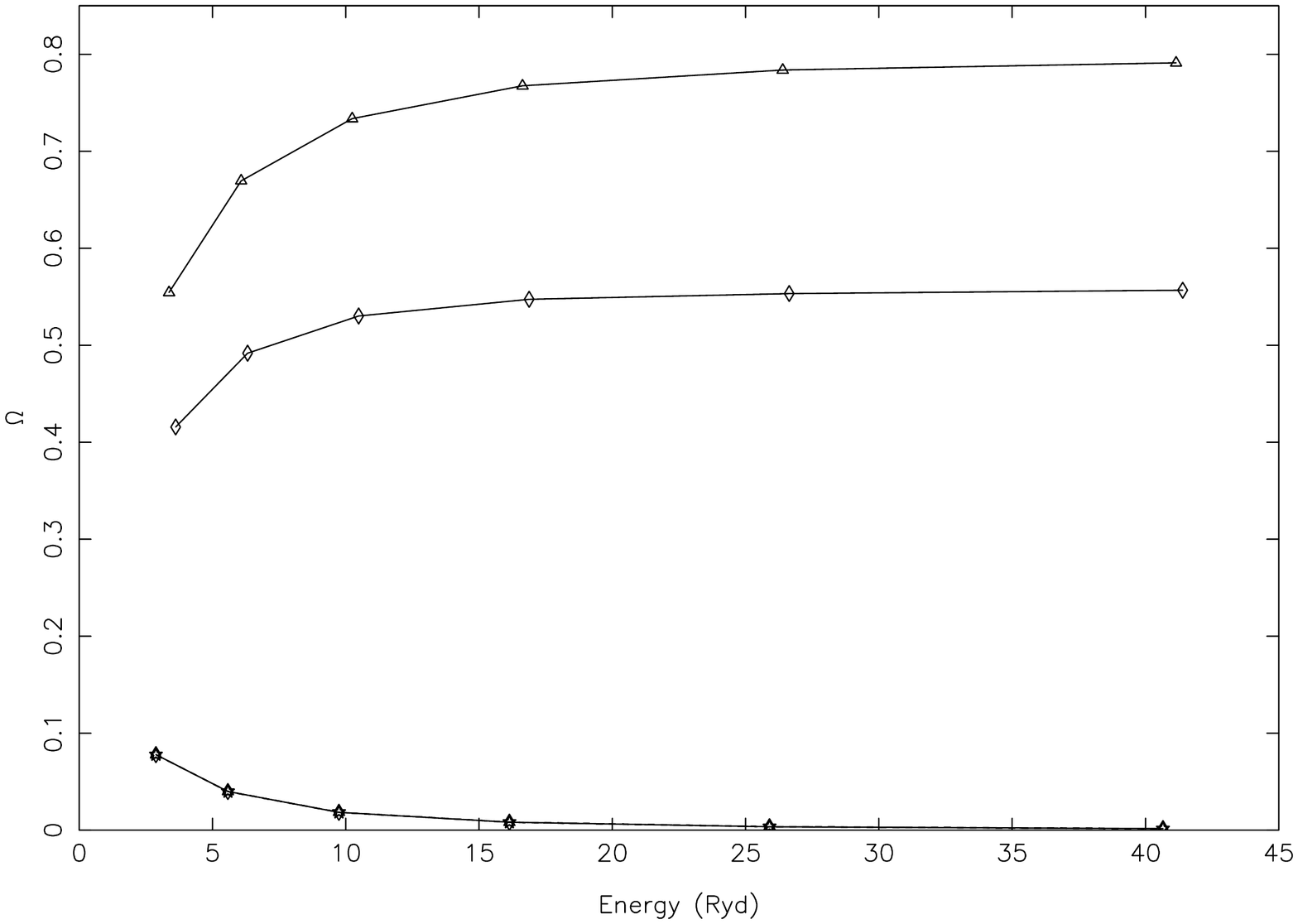}
 \vspace{-1.5cm}
\caption{Our calculated values of  $\Omega$   with FAC for  the 2--51 (triangles: 2p$^2$~$^3$P$_1$--2s2p$^2$($^4$P)3s~$^3$P$_1$), 2--72 (diamonds: 2p$^2$~$^3$P$_1$--2s$^2$2p($^2$P$^o$)4p~$^3$P$_1$), and 3--23 (stars: 2p$^2$~$^3$P$_2$--2s$^2$2p($^2$P$^o$)3p~$^1$P$_1$)  transitions of O~III. }\label{f9}

 \end{figure*}

\begin{figure*}
\setcounter{figure}{9}
\includegraphics[angle=-90,width=0.9\textwidth]{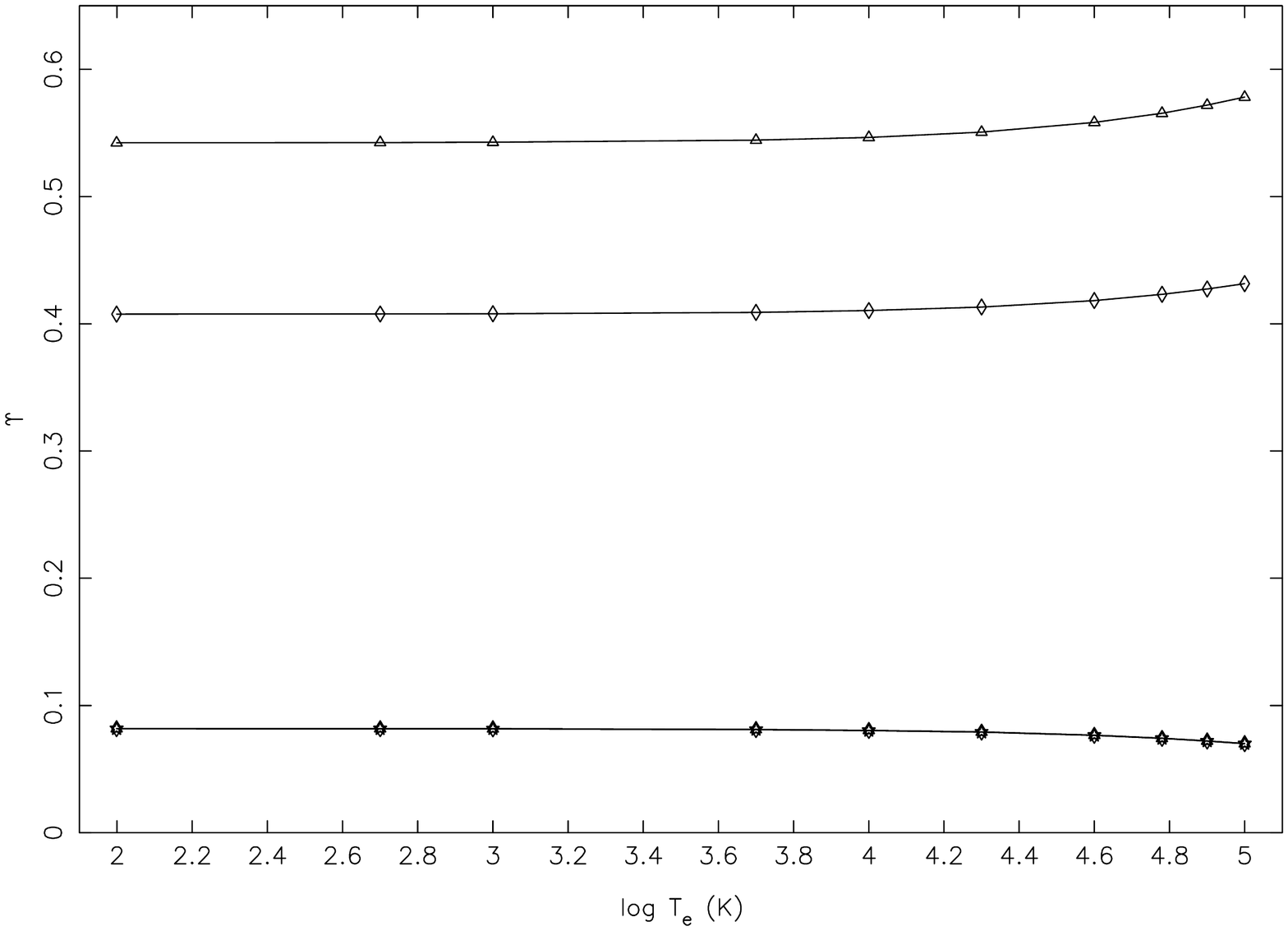}
 \vspace{-1.5cm}
\caption{Our calculated values of   $\Upsilon$  with FAC for  the 2--51 (triangles: 2p$^2$~$^3$P$_1$--2s2p$^2$($^4$P)3s~$^3$P$_1$), 2--72 (diamonds: 2p$^2$~$^3$P$_1$--2s$^2$2p($^2$P$^o$)4p~$^3$P$_1$), and 3--23 (stars: 2p$^2$~$^3$P$_2$--2s$^2$2p($^2$P$^o$)3p~$^1$P$_1$)  transitions of O~III. }\label{f10}

 \end{figure*}

\begin{figure*}
\setcounter{figure}{10}
\includegraphics[angle=-90,width=0.9\textwidth]{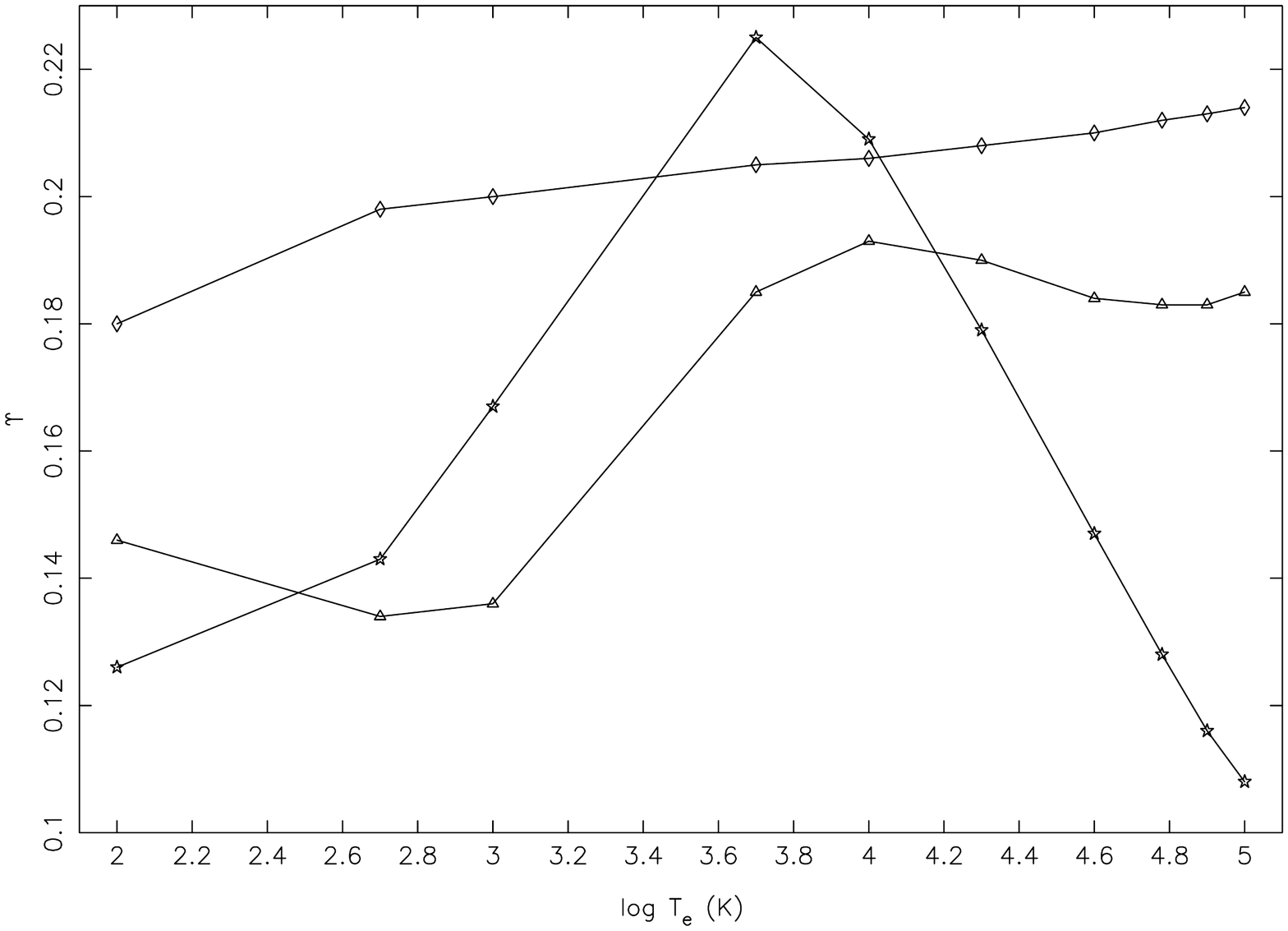}
 \vspace{-1.5cm}
 \caption{The  calculated values of  $\Upsilon$  by Tayal and Zatsarinny  \cite{sst} with BSR for  the 2--51 (triangles: 2p$^2$~$^3$P$_1$--2s2p$^2$($^4$P)3s~$^3$P$_1$), 2--72 (diamonds: 2p$^2$~$^3$P$_1$--2s$^2$2p($^2$P$^o$)4p~$^3$P$_1$), and 3--23 (stars: 2p$^2$~$^3$P$_2$--2s$^2$2p($^2$P$^o$)3p~$^1$P$_1$)  transitions of O~III. }\label{f11}
 \end{figure*}

Some differences in the $\Omega$, and subsequently $\Upsilon$ values, shown in Figure~\ref{f8} can be due to the differences in the corresponding f-values, because our wavefunctions are rather simple whereas those of Tayal and  Zatsarinny  \cite{sst} are more sophisticated with the inclusion of larger CI (configuration interaction). Therefore, as a test we have performed additional  calculations by including further CI with the (2p$^3$) 3$\ell$ and 4$\ell$ configurations. These seven configurations give rise to additional 192 levels (i.e. 394 in total) but all (except one) have energies higher than the 202 considered earlier. Effectively there is no close interaction between these additional levels/configurations and those considered earlier. For this reason the f-value for the 83--98 (2s$^2$2p($^2$P$^o$)4d~$^3$F$^o_4$--2s$^2$2p($^2$P$^o$)4f~$^3$G$_5$) transition is invariant ($\sim$0.14) and hence should not affect the calculations of $\Omega$ and $\Upsilon$. The other two transitions, namely 89--93 (2s$^2$2p($^2$P$^o$)4d~$^3$D$^o_3$--2s$^2$2p($^2$P$^o$)4f~$^3$F$_4$) and 94--101 (2s$^2$2p($^2$P$^o$)4d~$^1$F$^o_3$--2s$^2$2p($^2$P$^o$)4f~$^1$G$_4$), are comparatively weaker, and therefore for the former it changes from 0.030 to 0.022 with the inclusion of larger CI but not for the latter, which remains $\sim$0.016. This conclusion has been confirmed with the similar calculations performed with the GRASP (general-purpose relativistic atomic structure package) code.
 
 Before concluding we consider three more transitions, namely 2--51 (2p$^2$~$^3$P$_1$--2s2p$^2$($^4$P)3s~$^3$P$_1$), 2--72 (2p$^2$~$^3$P$_1$--2s$^2$2p($^2$P$^o$)4p~$^3$P$_1$), and 3--23 (2p$^2$~$^3$P$_2$--2s$^2$2p($^2$P$^o$)3p~$^1$P$_1$), all of which are forbidden. Our calculated values with FAC for $\Omega$ and $\Upsilon$ are shown in Figures~\ref{f9} and \ref{f10}, and the  results of Tayal and  Zatsarinny  \cite{sst} in Figure~\ref{f11}. Our values for both parameters vary smoothly, mainly because we have not considered the contribution of resonances, but Tayal and Zatsarinny have. However, not only the magnitudes of their $\Upsilon$ results are lower, the behaviours also appear to be abnormal, particularly of the 3--23 transition --  note the sharp fall in magnitude at temperatures above 10$^{3.7}$~K. 

\section{Conclusions}\label{sec3} 

In this paper we have made an assessment of the $\Upsilon$ results  reported by  Tayal and  Zatsarinny  \cite{sst} for transitions among 202 levels of O ~III. It is a pity that even after performing sophisticated calculations with elaborate wavefunctions (which involved up to 800 configurations for each level) and using large computational resources (which involved up to 256 processors handling Hamiltonian of the sizes as large as 70~000 for a partial wave), the reported results are found to be anomalous,  deficient and inaccurate for several transitions of all types, such as:  allowed, forbidden, intercombination, and semiforbidden, and at all temperatures. We have arrived at this conclusion based on comparisons with our own independent (non-resonant) calculations with FAC and our long experience with similar work on a wide range of ions. For the large discrepancies noted, in both magnitudes and behaviours, we have identified three possible sources of errors, which are: (i) inclusion of insufficient number of partial waves, (ii) a coarse energy mesh at energies above thresholds, and (iii) unrequired extrapolation of $\Omega$ to energies higher than actually calculated. 

Excitation rates for the transitions of O~III are required for up to (at least) 1.62$\times$10$^5$~K, the temperature derived by  Doschek \cite{gad} and Keenan and Aggarwal \cite{fpk}  from the measured line intensities in the solar transition regions. For this reason in our earlier work \cite{oiii} we reported values of $\Upsilon$ up to T$_e$ = 2.0$\times$10$^5$~K. Therefore, it is not understandable why Tayal and  Zatsarinny    \cite{sst}  restricted their results up to T$_e$ = 10$^5$~K, in spite of calculating $\Omega$ up to an energy of 30~Ryd, in comparison to only 13~Ryd by us \cite{oiii}. Perhaps they were aware of the inaccuracies at higher temperatures! Unfortunately, their listed results cannot be extrapolated to higher temperatures, because this will lead to much larger inaccuracies than noted already. Furthermore, for the diagnostics and modelling of plasmas, a {\em complete} set of atomic data for energy levels, A-values and $\Upsilon$ (and preferably for $\Omega$ too because these are very useful for accuracy assessments) are required, but Tayal and Zatsarinny  have reported A-values for only a few transitions, among the lowest 15 levels alone. 

We will like to stress here that the discrepancies in $\Upsilon$ values noted in this paper for several transitions of O~III, and earlier for other ions such as Mg~V \cite{mgv} and S~III \cite{s3}, are not because of the code (BSR) adopted by them, but its implementation. If the code is used judiciously with due care for the requirements than the results obtained should be in reasonable agreement with those from others, such as the standard $R$-matrix code, Dirac atomic $R$-matrix code (DARC), or even FAC, particularly for the allowed transitions. Therefore, in our opinion fresh calculations (with any version of the $R$-matrix code) should be performed for O~III, an ion of great astrophysical importance. And this is in spite of the availability of more recent and larger calculations by Mao et al. \cite{icft}, which have already been assessed to be inaccurate by Morisset et al. \cite{atoms}. Finally, our results with FAC can either be easily generated or  obtained from the author on request.


\end{document}